# Andean terraced hills (a use of satellite imagery)

**Amelia Carolina Sparavigna**
Dipartimento di Fisica, Politecnico di Torino
C.so Duca degli Abruzzi 24, Torino, Italy

**Abstract**
The aim of this paper is in stimulating the use of satellite imagery, in particular the free service of Google Maps, to investigate the distribution of the agricultural technique of terraced hills in Andean countries, near Titicaca Lake. In fact, satellite maps can give a clear view of the overall surface modified by human work, being then a precious help for on-site archaeological researches and for historical analysis. Satellite imagery is also able to give the distribution of burial and worship places. The paper discusses some examples near the Titicaca Lake.

In previous papers [1-3] we proposed the used the Google satellite imagery to investigate the "raised fields" in the region of Titicaca Lake. We clearly see these fields, which are pre-Incaic earthworks with different forms and size, generally being 4-10 m wide, 10 to 100 m long, in the plain regions near the lake. In spite of erosion, the raised fields are clearly visible from the space. The net of earthworks appears in the satellite imagery as clearly planned, following the natural slope of the terrain, with many canals and artificial ponds. Some specific points of the net have earthworks creating symbols that look as animals or have a geometric design. They can be considered as geoglyphs: these locations are currently under investigation by Peruvian archaeologists, that verified the presence of "waru-waru", as the raised fields are locally called [4].
Waru-waru were used to enhance the agricultural production, increasing moisture conditions for crops. Besides the raised fields, the region is covered by terraced hills [6-8]. Since these hills are in strongly connection with the presence of waru-waru, we could argue that these engineered hills are also of a product of a pre-Incaic period [7]. The remains of this overall agricultural system are providing evidence of the impressive engineering abilities of the peoples who lived in Andean countries many centuries before Columbus arrive.
Let us remember that Lake Titicaca sits 3,811 m above sea level, in a basin high in the Andes on the border of Peru and Bolivia. The western part of the lake lies within the Puno Region of Peru, and the eastern side is located in the Bolivian La Paz Department. Both regions have the slopes of the hills criss-crossed with terrace walls and the plains covered with raised fields. Of course, the terraced hills can clearly observed by means of Google satellite imagery. Unfortunately, the imagery has not everywhere a high resolution and then a detailed analysis of all the hills near Titicaca Lake is not possible. Nevertheless, we would point out that an interesting application of the use of satellite imagery could be a survey of the region to evaluate the total agricultural extension of raised fields and terraced hills, with related artificial canal and ponds.
Another interesting application of Google Maps is for an evaluation of the engineering ability of the people who built these structures. For instance, let us observe a terraced hill of Atuncalla, near the Titicaca Lake. Figure 1 is obtained with Google: we see the almost perfect bell-shape of this hill, with lines, that seems to joint points of equal level, as in cartography the contour lines. We see also lines that are perpendicular to them, that is lines following the gradient. Let us remember that in the vector calculus, the gradient of a scalar field is a vector field which points in the direction of the greatest rate of increase of the scalar field, and whose magnitude is the greatest rate of change. In the case of Figure 1, the scalar function is the height on the sea-level and the gradient is the highest slope direction. This means that people arranged the walls to have the highest effect in reducing the erosion of soil by atmospheric actions.

Increasing the resolution, we see objects that seems to be heaps or piles of stones, placed almost at the center of terraced fields (see Figure 1, low panel). On the top of another hill, we see two of these heaps (see Figure 2): increasing the resolution and enhancing the images with a wavelet filter [9], one of this objects seems to have a circular hollow shape. This object could be a chullpa, that is a burial tower of pre-Incaic people. Similar objects can be observed on the terraced fields of the hills in the Bolivian region near the Titicaca Lake. Figure 3 shows a place in Bolivia that was localized in Ref.10 (the authors of this web page are using three-dimensional rendering of Google Maps to illustrate the terraced hills).

To understand if these objects can be burial or ancestor worship places, let us consider a well-know archaeological site near Titicaca Lake. This place is Sillustani, the peninsula of Umayo Lagoon. At the top of the peninsula there is a tall tower, about 12m high. According to scholars, Sillustani is a pre-Incan burial area: tombs are built above ground in the chullpas tower-like structures. These structures were built by Colla people, Aymara, who were conquered by the Inca in the 15th century. The structures housed the remains of complete family groups, although they were probably limited to nobility [11]. Wikipedia is reporting several interesting information on chullpas, that the ancestor worship and kinship were integral parts of Aymara culture and that the insides of tombs were shaped like a woman's uterus, and corpses were mummified in a fetal position to recreate their birth. The only openings to the buildings face east, where it was believed the Sun was reborn by Mother Earth each day. Another interesting remarks is on the architecture of chullpas. Wikipedia is also telling that "the architecture of the site (Sillustani) is often considered more complex than typical Incan architecture. In contrast with the Inca, who used stones of varying shapes, the Colla used even rectangular edges. While chullpas are not unique to Sillustani and are found across the Altiplano, this site is considered the best and most preserved example of them". Figure 4 shows as Sillustani can be observed with Google Maps. The upper image is that obtained from ACME mapper, the lower the same images after a processing with a wavelet filtering.

From the comparison of Figures 1-3 and 4 obtained by Google Maps, we can argue that the terraces hills could be sometimes burial or worship places. This means then a strong connection between living people and their ancestors. In fact, from a web site [12], we can understand that the "current theory about Andean mortuary practices focuses on the connections between the living and the deceased, and their relationship to the landscape and built environment. Ancestor worship permeated all levels of Andean society from the local community to the Inca state. Andean people venerated and buried their deceased in many ways. In the altiplano of Bolivia, the Aymara people placed some members of their society in … chullpas. Throughout the Sajama region, lines or ritual pathways are etched into the landscape. The ethnographies of the area suggest a link between the lines and some cultural features… The research will also address other possible associations between chullpas and the cultural landscape."

It seems then that a use of Google Maps (or other satellite services) devoted to the analysis of "landscapes and built environments", can be of great help for archaeological and historical researches, in particular for Andean countries, where many locations seems to be well-preserved and not revolutionized by modern cultivation techniques. Once more, the public satellite maps service shows itself as a powerful tool for the analysis of the cultural and historical landscape.

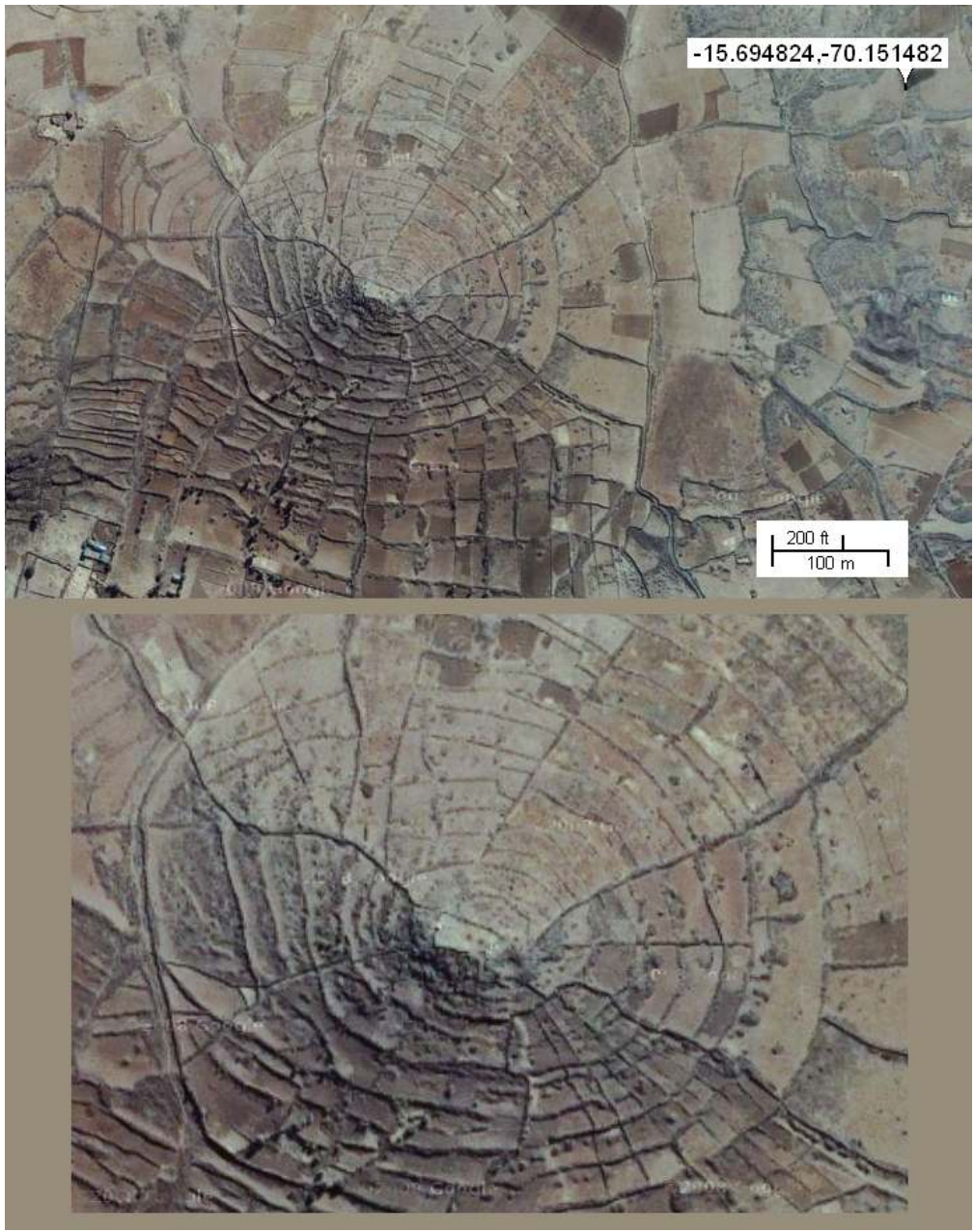

Figure 1: An image obtained with Google Maps of a hill in Antucalla near the Titicaca Lake. We see the bell-shaped hill, with lines, that seems to joint points of equal level, as in cartography the contour lines. There are also perpendicular lines that seems to follow the gradient. Let us remember that the gradient of a scalar field is a vector field which pointing in the direction of the greatest rate of increase of the scalar field. In this case, the scalar function is the height on the sea-level and the gradient is the highest slope direction. The engineered structure of the walls had the effect to reduce the erosion of soil.

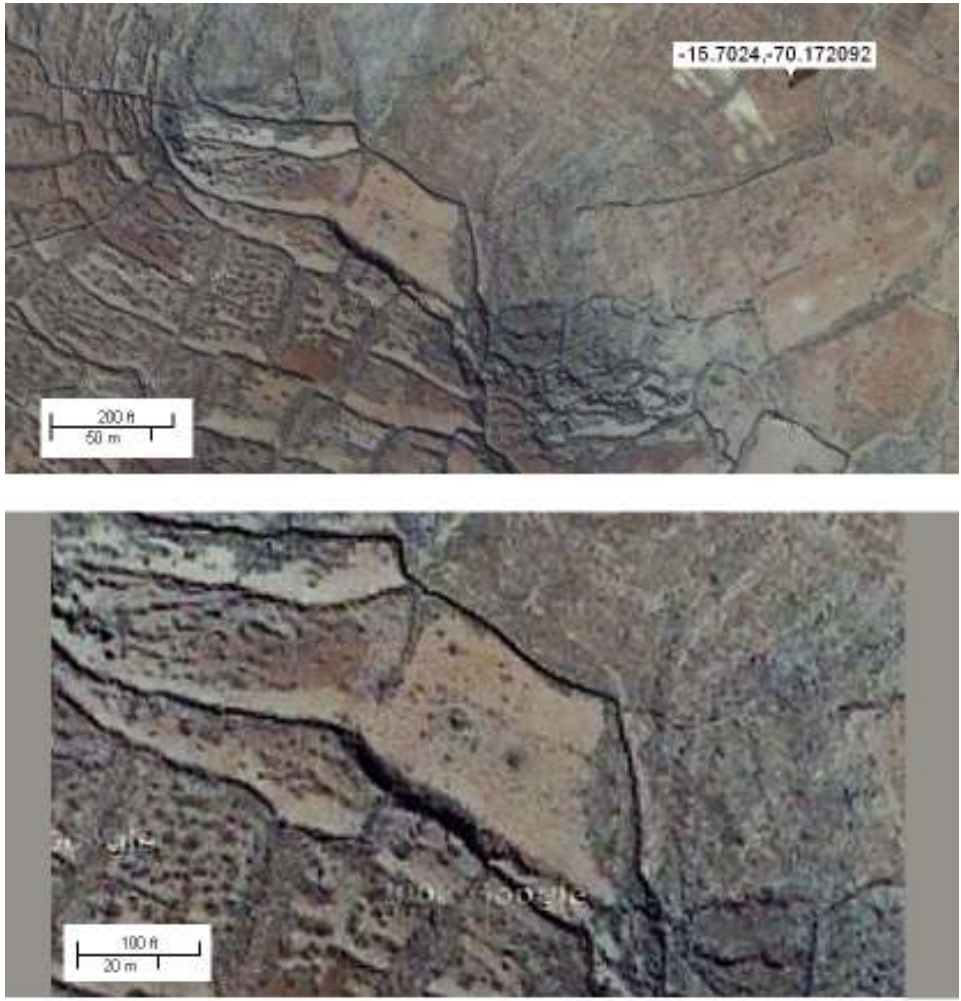

Figure 2: Another hill of Atuncalla. Increasing the resolution, we see objects that seems to be heaps of stones, places almost at the center of terraced fields. On the top of the hill, we see two of these heaps: increasing the resolution and enhancing the images with a wavelet filter (lower panel), one of this objects seems to have a circular hollow shape. Is this a burial tower or any other worship place?

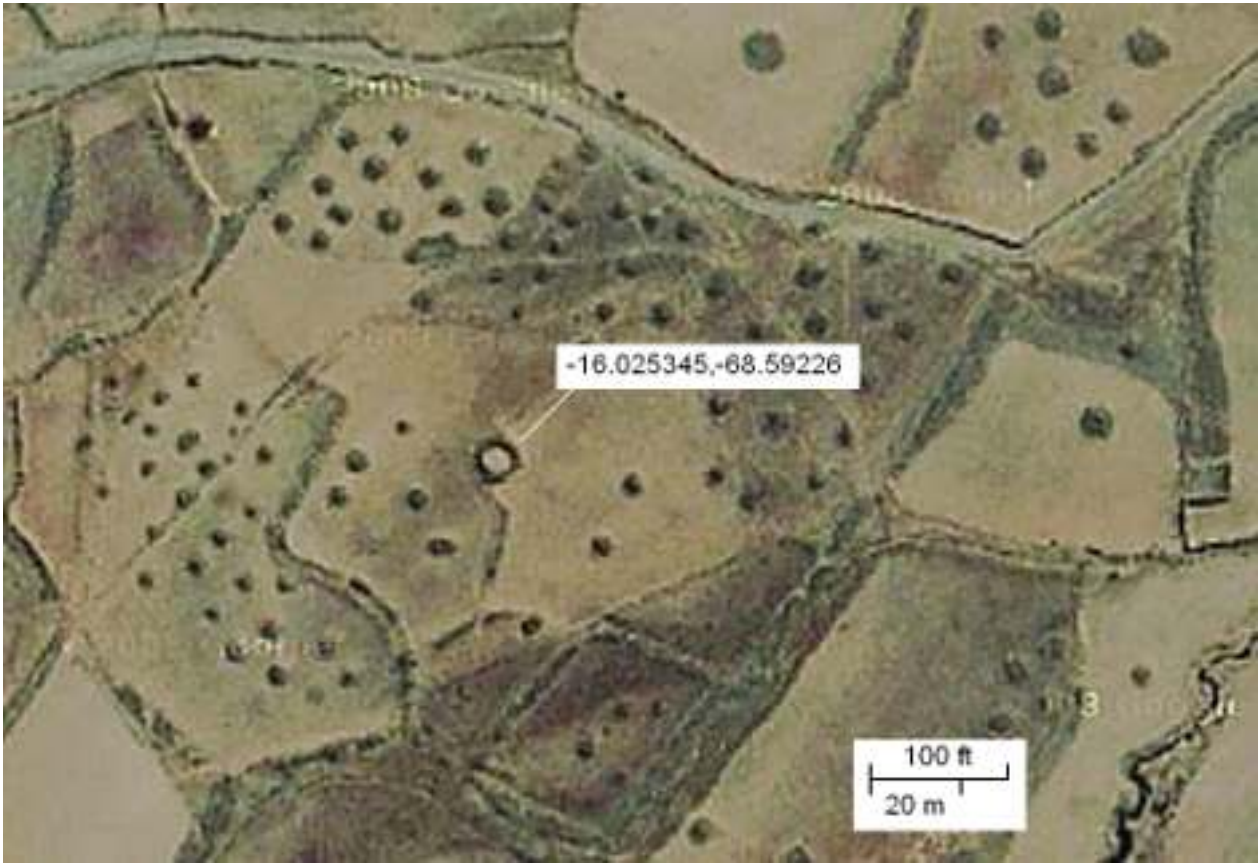

Figure 3: A hill near Titicaca Lake in Bolivia [localized by Ref.10 in three-dimensional rendering of Google Maps]. Is the circular object and the near fields the remains of an ancestor worship place?

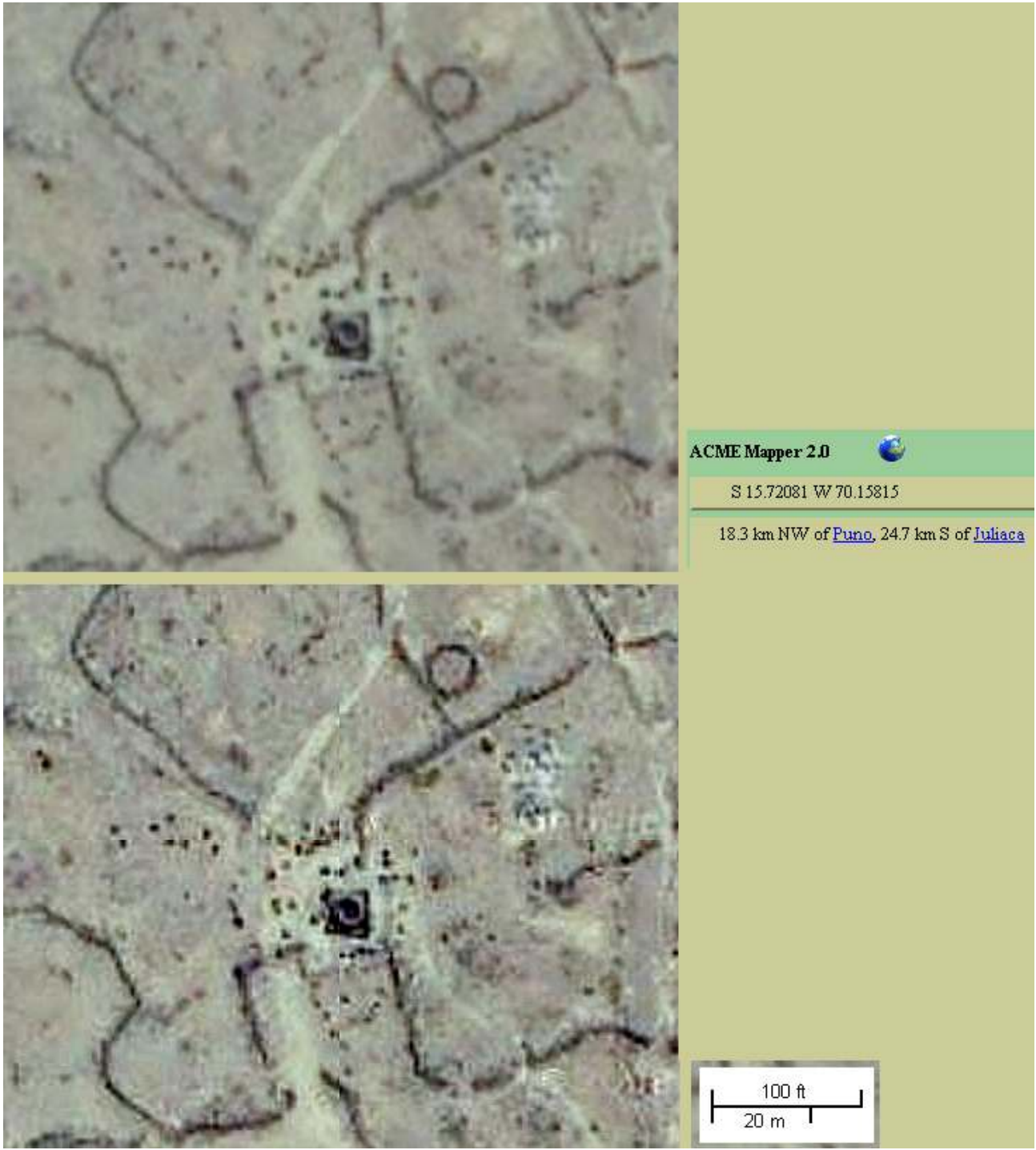

Figure 4: The top of the Sillustani peninsula as its appears from Google Maps. The upper image is as it is obtained from ACME mapper, a Google service. The lower panel shows the same images after a processing with a wavelet filtering. Sillustani is an example of burial and ancestor worship place. In the image, it is possible to see a well-preserved tower-like structure, called chullpa.